# Physics-guided residual Kalman learning for state-of-charge estimation of lithium iron phosphate batteries

Feng Guo[a,b,c], Luis D. Couto[a,b], Khiem Trad[a,b], Ru Hong[d,*], Guangdi Hu[e,*], Mohammadhosein Safari[b,c,f,*]

[a]WET, VITO, Boeretang 200, Mol, 2400, Belgium
[b]EnergyVille, Thor Park 8310, Genk, 3600, Belgium
[c]Institute for Materials Research (IUMAT), Hasselt University, Martelarenlaan 42, Hasselt, 3500, Belgium
[d]College of Tianyin Lake Science and Technology Innovation, Nanjing Institute of Technology, Nanjing 211167, Jiangsu, China
[e]School of Vehicles and Intelligent Transportation, Fuyao University of Science and Technology, Fuzhou 350109, Fujian, China
[f]IUMAT, IMEC, Diepenbeek, 3590, Belgium

**[*]Corresponding author.**
E-mail addresses: momo.safari@uhasselt.be (M. Safari), ruhong_ai@163.com (R. Hong), guangdihu@fyust.edu.cn (G. Hu).

## ABSTRACT

Accurate state of charge (SOC) estimation of lithium iron phosphate (LFP) batteries remains challenging because of their flat open-circuit-voltage (OCV)-SOC characteristics, temperature-dependent dynamics, and sensitivity to initialization errors. Here, we propose a physics-guided residual Kalman learning (PRKL) framework for electrochemical-model-based SOC estimation. PRKL combines a control-oriented single-particle-model-based extended Kalman filter (EKF), which provides recursive physical state propagation, with a gated recurrent unit (GRU) residual learner that compensates structured EKF errors using electrochemical states and measurement features. The framework is evaluated on a public graphite/LFP dataset covering three dynamic drive cycles, eight temperatures from −10 to 50 °C, and initialization offsets up to 20%. Using dynamic stress test (DST) and federal urban driving schedule (FUDS) cycles for training and the supplemental federal test procedure (US06) cycle for cross-profile testing within the same cell dataset, PRKL achieves a global average root mean square error (RMSE) of 1.19%, corresponding to a 77% reduction relative to the physics-only EKF. These results show that electrochemical state information can guide residual learning and improve recursive SOC estimation for LFP batteries. The present validation supports cross-profile robustness within the studied dataset and provides a basis for future cross-cell, ageing-aware, and embedded-platform validation.

## 1 Introduction

Lithium iron phosphate (LFP) batteries, owing to their superior safety and lower cost compared to nickel-manganese-cobalt (NMC) counterparts [1–4], have been widely deployed in electric transportation [5,6] and stationary energy storage applications [7]. Accurate estimation of the internal battery states, particularly the state of charge (SOC) that reflects the remaining capacity, is crucial for ensuring safe and efficient operation [8–10]. However, the dominant two-phase Li insertion mechanism of the LFP electrode yields an exceptionally flat open-circuit voltage (OCV)-SOC profile, posing a fundamental challenge for reliable SOC estimation in battery management systems (BMS) [11–13]. This challenge becomes even more pronounced under dynamic load profiles and wide temperature variations, where model uncertainties and initial SOC errors can further deteriorate the estimation accuracy.

One of the primary obstacles to high-accuracy SOC estimation for LFP is the limited fidelity of battery models. Because of the flat OCV-SOC relationship, even small model errors can introduce significant noise, hindering the ability of algorithms to deliver reliable estimates. The equivalent circuit model (ECM) is the most widely used in practice, where voltage sources, resistors, and capacitors are employed to mimic the battery terminal behavior [14,15]. While computationally simple, ECMs cannot capture the underlying physicochemical processes and often suffer from large modeling errors. In contrast, electrochemical models, such as the pseudo-two-dimensional (P2D) model [16] and its simplified variants, explicitly describe internal electrochemical reactions and have therefore attracted increasing attention for state estimation [8]. Nevertheless, their reliance on solving coupled partial differential equations leads to a high computational cost, which limits their real-time applicability [17]. To balance physical interpretability and computational efficiency, the single particle model (SPM) [18] has emerged as a practical alternative, as it preserves the essential electrochemical features while simplifying the model structure and computation, making it suitable for deployment in BMS.

Model-based approaches for battery SOC estimation mainly include Kalman filtering [19], particle filtering [20], sliding mode observers [21,22], $H_\infty$ observers [23], and among others. These methods largely rely on an underlying battery model, the fidelity of which directly influences the estimation accuracy of the method. The performance of a filter, even with more accurate electrochemical models, will remain vulnerable to errors induced by the model simplifications and uncertainties in the parameterization. As a result, machine learning-based SOC estimation methods have been widely adopted as an alternative [24]. These approaches typically predict the SOC directly from measurement data without relying on explicit battery models. Representative techniques include long short-term memory (LSTM) networks [25], transformer [26], temporal convolutional network (TCN) [27], convolutional neural network (CNN) [28], and other deep learning architectures [29–31]. However, purely data-driven models lack physical constraints, making them prone to overfitting and limited generalization. In addition, their poor interpretability hinders fault diagnosis in safety-critical applications [32]. This motivates the integration of physics-based models with machine learning to enhance both accuracy and interpretability [33].

Current hybrid approaches that combine machine learning and physics-based models for battery SOC estimation can be broadly categorized into three types. The first category leverages machine learning to improve model-based filters [34]. For example, Zhao et al. employed KalmanNet based on an ECM, where a gated recurrent unit (GRU) was used to learn the Kalman gain to enhance the accuracy of Kalman filtering [35]. Similarly, Rezaei et al. incorporated a fuzzy system to adaptively adjust the measurement noise covariance of the Kalman filter, which also led to a modest performance gain [36]. However, the improvement was limited by the accuracy of the ECM itself and the inherent structure of the Kalman filter. The second category enhances the accuracy of battery models using machine learning. For instance, Feng et al. integrated the SPM with a neural network to refine the electrochemical model, followed by SOC estimation using an unscented Kalman filter (UKF) [37]. Likewise, Zheng et al. employed physics-informed neural networks (PINNs) to solve the partial differential equations of the electrochemical model before applying UKF for SOC estimation [38]. While these methods improve model fidelity, their final performance still remains constrained by the structure of the Kalman filter. The third category focuses on fusing the outputs of model-based and machine learning methods [39]. For example, Zeng et al. estimated SOC separately using a UKF and an LSTM network, and then employed another LSTM to refine the final output [40]. This approach suffers from high computational complexity, and the machine learning component still lacks physical information.

For SOC estimation of LFP batteries, Tian et al. employed a deep neural network (DNN) to replace the battery model and fused its output with Coulomb counting through a Kalman filter, achieving an error of 2.03% [41]. However, their approach was only validated under constant-current discharge conditions, without considering dynamic operating profiles or evaluating robustness against initialization errors. Moreover, this framework lacked a temperature model, limiting its applicability to realistic environmental conditions. Che et al. proposed a model fusion strategy that combined random forests, Coulomb counting, and the open-circuit voltage method, and demonstrated SOC estimation under dynamic operating conditions and temperatures ranging from 10 to 45 °C, with a mean absolute error (MAE) of 3.25% [11]. Hou et al. designed an extended Kalman filter (EKF) based on an ECM and further employed an XGBoost model to correct SOC estimation [42]. Despite these advances, existing approaches typically fall into two categories: either physics-based filters constrained by simplified equivalent circuit representations, or purely data-driven predictors lacking mechanistic interpretability. In both cases, the estimation framework remains tightly coupled to either model fidelity or data representational power. As a result, robust high-accuracy SOC estimation under dynamic load profiles, wide temperature ranges, and significant initialization errors remains unresolved for LFP systems.

In this work, we propose a physics-guided residual Kalman learning (PRKL) framework that decomposes hybrid SOC estimation into two coupled but separable channels: a physics-based recursive estimation channel and a data-driven residual correction channel. A control-oriented electrochemical single-particle model with thermal effects is embedded in an extended Kalman filter (EKF) to provide physically meaningful state propagation and a baseline SOC estimate, while a GRU-based residual learner operates in parallel to compensate structured model discrepancies. Unlike hybrid strategies that replace the physical model or directly learn the Kalman gain, PRKL preserves the electrochemical EKF as the primary estimator and uses learning only as an explicit residual correction layer. In this sense, the method is physics-guided through EKF-derived electrochemical states and measurement innovations, rather than being a fully physics-constrained neural network with governing-equation loss terms [43]. The contribution of this study is therefore a mechanism-level demonstration that electrochemical state information can substantially improve residual learning for LFP SOC estimation within a recursive estimation framework.

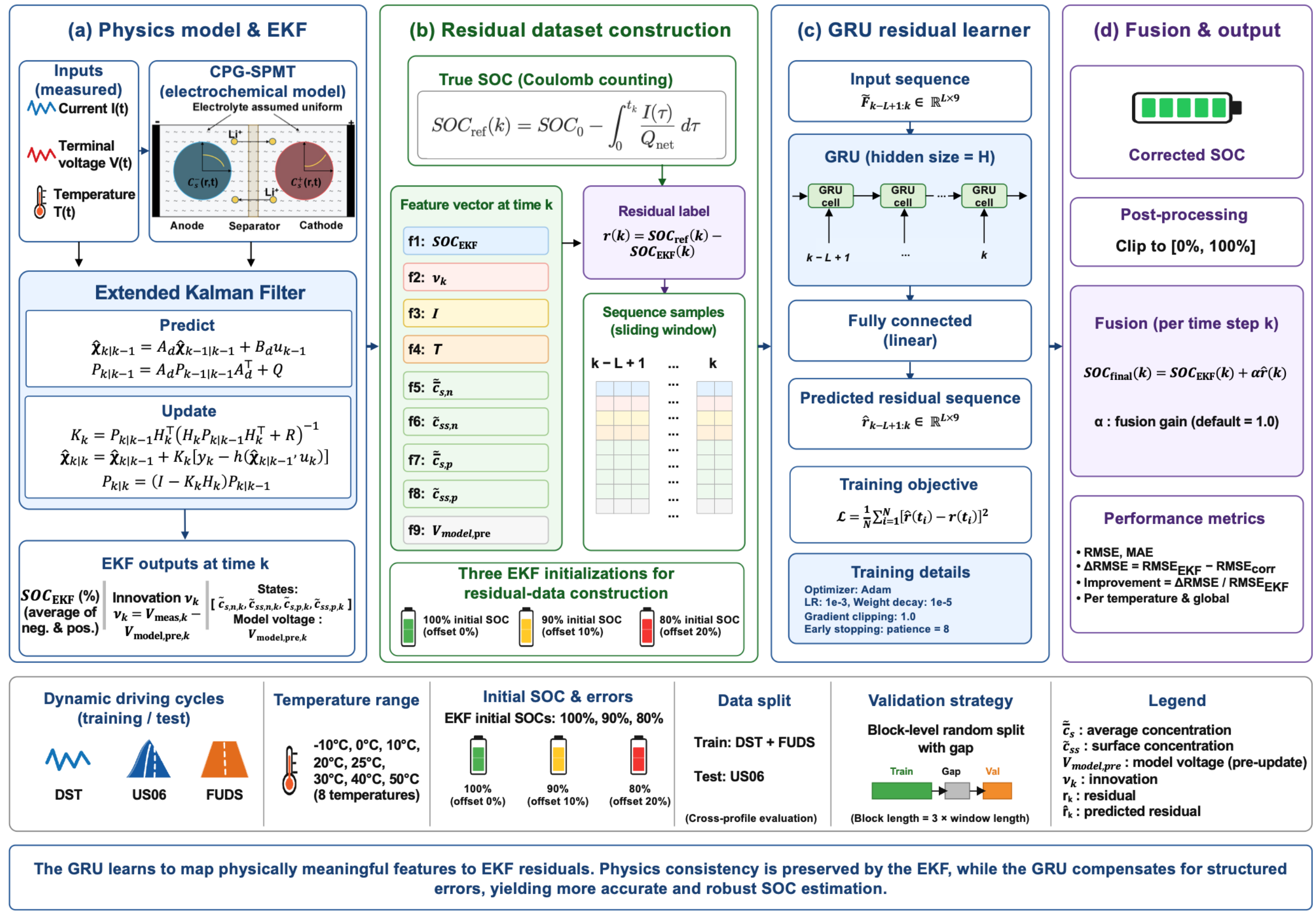

**Fig. 1.** Overview of the proposed PRKL framework for SOC estimation of LFP battery. (a) The control-oriented parameter-grouped single particle model (CPG-SPMT) is embedded in an EKF to generate the baseline SOC estimate, voltage innovation, pre-update model voltage, and electrode-level states from measured current, voltage, and temperature. (b) Residual learning data are constructed using Coulomb-counting reference SOC, EKF residuals, nine input features, and three EKF initialization settings. (c) A GRU maps windowed feature sequences to SOC residuals. (d) The corrected SOC is obtained by fusing the EKF estimate with the GRU-predicted residual and clipping the output to 0%–100%. The bottom panel summarizes the validation settings.

## 2 Methods

### *2.1 Framework of PRKL*

The overall framework of the proposed PRKL method is illustrated in Fig. 1. PRKL combines a physics-based recursive estimation channel with a data-driven residual correction channel. As shown in Fig. 1(a), the measured current, terminal voltage, and temperature are first processed by the control-oriented parameter-grouped single particle model with thermal effects (CPG-SPMT) embedded in an EKF. The EKF recursively propagates the internal electrochemical states and updates the baseline

SOC estimate using voltage measurements. Its outputs include the EKF-estimated SOC, voltage innovation, pre-update model voltage, and electrode-level average and surface concentration states.

Fig. 1(b) shows the construction of the residual learning dataset. The reference SOC is obtained by Coulomb counting, and the learning target is defined as the residual between the reference SOC and the EKF-estimated SOC. For each time step, a nine-dimensional feature vector is constructed from EKF outputs, measurement variables, and electrochemical state information. Three EKF initial SOC settings, namely 100%, 90%, and 80%, are used to construct residual data under initialization offsets of 0%, 10%, and 20%, respectively. Together with three dynamic load profiles and eight temperatures from −10 to 50 °C, this design enables the residual learner to capture structured estimation errors under diverse operating conditions.

As shown in Fig. 1(c), the GRU residual learner takes temporally windowed feature sequences as inputs and predicts the SOC residual sequence. Unlike purely data-driven SOC estimators, the GRU does not replace the electrochemical estimator; instead, it learns the systematic residual errors left by the reduced-order electrochemical EKF. Finally, as illustrated in Fig. 1(d), the corrected SOC is obtained by fusing the EKF estimate with the GRU-predicted residual and clipping the result to the physically meaningful range of 0%–100%.

This design should be interpreted as physics-guided residual learning rather than a fully physics-informed neural network. Charge conservation and electrochemical dynamics are enforced in the CPG-SPMT-EKF propagation channel, whereas the GRU serves as a lightweight residual corrector guided by physically meaningful state features. Thus, the neural network compensates for structured model and filter errors while preserving the interpretability and recursive structure of the electrochemical estimator.

### *2.2 Electrochemical model and EKF*

We employ a CPG-SPMT as the electrochemical backbone for state estimation [18,43]. Starting from the classical SPM, the solid-phase diffusion in representative spherical particles is described by Fick's law, and the terminal voltage is computed from open-circuit potentials, charge-transfer overpotentials, and ohmic loss. The full SPM formulation is summarized in Table 1.

**Table 1.** Original continuous-time SPM.

| Component | Governing equation |
|---|---|
| Solid-phase diffusion | $\dfrac{\partial c_{\mathrm{s},i}}{\partial t} = \dfrac{1}{r^2}\dfrac{\partial}{\partial r}\left(D_{\mathrm{s},i} r^2 \dfrac{\partial c_{\mathrm{s},i}}{\partial r}\right), \quad i \in \{n, p\}$ |
| Initial condition | $c_{\mathrm{s},i}(r, 0) = c_{i,0}$ |
| Center boundary | $\left.\dfrac{\partial c_{\mathrm{s},i}}{\partial r}\right\|_{r=0} = 0$ |
| Surface boundary | $\left.\dfrac{\partial c_{\mathrm{s},i}}{\partial r}\right\|_{r=R_{\mathrm{s},i}} = -\dfrac{j_i(t)}{D_{\mathrm{s},i}}$ |
| Surface normalization | $\tilde{c}_{\mathrm{ss},i} = \dfrac{c_{\mathrm{ss},i}}{c_{\max,i}}$ |
| Interfacial flux | $j_i(t) = 1_{\mp} \dfrac{I(t)}{F a_i A_i L_i}$ |
| Exchange current density | $j_{0,i} = r_{\mathrm{eff},i} c_{\max,i} \sqrt{c_{\mathrm{e}} \tilde{c}_{\mathrm{ss},i}(1 - \tilde{c}_{\mathrm{ss},i})}$ |
| Overpotential | $\eta_i = \dfrac{2RT}{F} \sinh^{-1}\left(\dfrac{1_{\mp} I}{2 a_i L_i j_{0,i}}\right)$ |
| Terminal voltage | $V_{\mathrm{SPM}} = \mathrm{OCP}_p(\tilde{c}_{\mathrm{ss},p}) - \mathrm{OCP}_n(\tilde{c}_{\mathrm{ss},n}) + \eta_p - \eta_n - R_{\mathrm{ini}} I$ |

To enable real-time filtering, the diffusion Partial Differential Equations（PDEs）are reduced via parabolic approximation and re-parameterized by grouping non-identifiable parameter products [18,43]. The resulting CPG-SPMT retains only two internal diffusion states per electrode and is summarized in Table 2. Specifically, the continuous-time state-space form is

$$\dot{\chi}(t) = \tilde{A}\chi(t) + \tilde{B}u(t), \quad \chi(0) = \chi_0 \qquad (1)$$

$$\psi(t) = \tilde{C}\chi(t) + \tilde{D}u(t) \qquad (2)$$

where $u(t) = I(t)$ is the applied current, $\chi$ stacks the diffusion states of both electrodes, and $\psi$ collects the normalized average and surface concentrations used by the nonlinear voltage function.

**Table 2.** Discretized and parameter-grouped control-oriented SPM (CPG-SPMT).

| **Item** | **Expression** |
|---|---|
| State-space form | $\dot{\chi}(t) = \tilde{A}\chi(t) + \tilde{B}u(t), \qquad \psi(t) = \tilde{C}\chi(t) + \tilde{D}u(t)$ |
| State vector | $\chi = \left[\chi_p^{\top} \quad \chi_n^{\top}\right]^{\top}, \quad \chi_i = [\tilde{q}_{1,i} \quad \tilde{q}_{2,i}]^{\top}, \quad i \in \{p, n\}$ |
| Output vector | $\psi = \left[\psi_p^{\top} \quad \psi_n^{\top}\right]^{\top}, \quad \psi_i = \left[\tilde{\bar{c}}_{\mathrm{s},i} \quad \tilde{c}_{\mathrm{ss},i}\right]^{\top}$ |
| Grouped matrices | $\tilde{A}_i = \begin{bmatrix} 0 & 0 \\ \dfrac{30}{\alpha_i} & -\dfrac{30}{\alpha_i} \end{bmatrix}, \quad \tilde{B}_i = \begin{bmatrix} \dfrac{1}{b_i} \\ \dfrac{19}{7b_i} \end{bmatrix}, \quad \tilde{C}_i = I_2, \quad \tilde{D}_i = \begin{bmatrix} 0 \\ \dfrac{\alpha_i}{105 b_i} \end{bmatrix}$ |
| Grouped parameters | $\alpha_i = \dfrac{R_{\mathrm{s},i}^2}{D_{\mathrm{s},i}}, \qquad b_i = F A_i L_i \varepsilon_i c_{\mathrm{max},i}$ |
| Grouped overpotential | $\eta_i = \dfrac{2RT}{F} \sinh^{-1}\left(\dfrac{1 \mp I}{6 b_i d_i \sqrt{\tilde{c}_{\mathrm{ss},i}\left(1 - \tilde{c}_{\mathrm{ss},i}\right)}}\right)$ |
| Reaction parameter | $d_i = \dfrac{r_{\mathrm{eff},i}\sqrt{c_{\mathrm{e}}}}{F R_{\mathrm{s},i}}$ |

Key transport/kinetic parameters are made temperature-dependent using Arrhenius-type relations to cover −10 to 50 °C, summarized in Table 3. Constants such as $R$, $F$, and $c_{\mathrm{e}}$ are treated as known and are not identified. Further details of the model parameters and the parameter identification procedure are provided in Ref. [44].

**Table 3.** Model parameters and temperature dependence.

| Parameter | Description | Temperature dependence |
|---|---|---|
| $b_n, b_p$ | Electrode capacities (grouped) | Constant |
| $\alpha_{n,1}, \alpha_{p,1}$ | Reference diffusion-group parameters | $\alpha_n(T) = \dfrac{\alpha_{n,1}}{\exp\left(\frac{E_1}{R}\left(\frac{1}{T_{\mathrm{ref}}} - \frac{1}{T}\right)\right)}$ <br> $\alpha_p(T) = \dfrac{\alpha_{p,1}}{\exp\left(\frac{E_2}{R}\left(\frac{1}{T_{\mathrm{ref}}} - \frac{1}{T}\right)\right)}$ |
| $d_{n,1}, d_{p,1}$ | Reference reaction-group parameters | $d_n(T) = d_{n,1}\exp\left(\frac{E_3}{R}\left(\frac{1}{T_{\mathrm{ref}}} - \frac{1}{T}\right)\right)$ <br> $d_p(T) = d_{p,1}\exp\left(\frac{E_4}{R}\left(\frac{1}{T_{\mathrm{ref}}} - \frac{1}{T}\right)\right)$ |
| $R_{\mathrm{ini},1}$ | Reference ohmic resistance | $R_{\mathrm{ini}}(T) = \dfrac{R_{\mathrm{ini},1}}{\exp\left(\frac{E_5}{R}\left(\frac{1}{T_{\mathrm{ref}}} - \frac{1}{T}\right)\right)}$ |
| $E_1, \dots, E_5$ | Activation energies | Constant |
| $\mathrm{SOC}_n(0), \mathrm{SOC}_p(0)$ | Initial electrode SOCs | Estimated |

With zero-order hold (ZOH) at $\Delta t$ = 1 s, the reduced model is discretized as

$$\chi_k = A_{\mathrm{d}}\chi_{k-1} + B_{\mathrm{d}}u_{k-1} \tag{3}$$

$$y_k = h(\chi_k, u_k) + v_k \tag{4}$$

where $A_{\mathrm{d}}$ and $B_{\mathrm{d}}$ are the discrete-time system matrices, $y_{\mathrm{k}}$ is the measured terminal voltage, and $v_{\mathrm{k}}$ denotes measurement noise. The nonlinear measurement function $h(\cdot)$ implements the CPG-SPMT voltage mapping using $\psi_k = C_{\mathrm{d}}\chi_k + D_{\mathrm{d}}u_k$.

The EKF estimates $\chi_k$ recursively via

$$\hat{\chi}_{k|k-1} = A_{\mathrm{d}}\hat{\chi}_{k-1|k-1} + B_{\mathrm{d}}u_{k-1} \tag{5}$$

$$P_{k|k-1} = A_{\mathrm{d}}P_{k-1|k-1}A_{\mathrm{d}}^{\top} + Q \tag{6}$$

$$K_k = P_{k|k-1}H_k^{\top}(H_kP_{k|k-1}H_k^{\top} + R)^{-1} \tag{7}$$

$$\hat{\chi}_{k|k} = \hat{\chi}_{k|k-1} + K_k[y_k - h(\hat{\chi}_{k|k-1}, u_k)] \tag{8}$$

$$P_{k|k} = (I - K_k H_k) P_{k|k-1} \tag{9}$$

where $H_k = \left.\frac{\partial h}{\partial \chi}\right|_{\hat{\chi}_{k|k-1}}$, and $Q$ and $R$ are the process and measurement noise covariance matrices.

From $\hat{\chi}_k$, the electrode-level outputs are reconstructed as $\psi_i = [\bar{\tilde{c}}_{\mathrm{s},i} \quad \tilde{c}_{\mathrm{ss},i}]^{\mathrm{T}}$ $(i \in \{p, n\})$. The normalized average concentrations are then mapped to electrode SOCs

$$\mathrm{SOC}_p(t) = \frac{\bar{\tilde{c}}_{\mathrm{s},p}(t) - \bar{\tilde{c}}_{\mathrm{s},p}^{\min}}{\bar{\tilde{c}}_{\mathrm{s},p}^{\max} - \bar{\tilde{c}}_{\mathrm{s},p}^{\min}} \tag{10}$$

$$\mathrm{SOC}_n(t) = \frac{\bar{\tilde{c}}_{\mathrm{s},n}(t) - \bar{\tilde{c}}_{\mathrm{s},n}^{\min}}{\bar{\tilde{c}}_{\mathrm{s},n}^{\max} - \bar{\tilde{c}}_{\mathrm{s},n}^{\min}} \tag{11}$$

and the overall cell SOC is defined as

$$\mathrm{SOC}_{\mathrm{EKF}}(t) = \frac{1}{2}\left[\mathrm{SOC}_p(t) + \mathrm{SOC}_n(t)\right] \tag{12}$$

*2.3 Physics-guided residual learning and fusion*

To further improve the SOC estimate provided by the model-based EKF, a GRU-based residual learner is introduced as a parallel correction module. Rather than predicting SOC directly from measurement sequences, the GRU learns the systematic residual of the EKF by using both measurable operating variables and EKF-derived electrochemical states. This design preserves the recursive electrochemical estimation channel while allowing the data-driven module to compensate for structured model mismatch and unmodelled nonlinearities.

The nine input features of the GRU residual model are summarized in Table 4. At each sampling instant *k*, the input vector is defined as

$$\mathrm{z}_k = [\bar{\tilde{c}}_{\mathrm{s},n,k}, \tilde{c}_{\mathrm{ss},n,k}, \bar{\tilde{c}}_{\mathrm{s},p,k}, \tilde{c}_{\mathrm{ss},p,k}, \mathrm{SOC}_{\mathrm{EKF},k}, V_{\mathrm{model,pre},k}, \nu_k, I_k, T_k]^{\mathrm{T}} \tag{13}$$

where $\bar{\tilde{c}}_{\mathrm{s},n,k}$, $\tilde{c}_{\mathrm{ss},n,k}$, $\bar{\tilde{c}}_{\mathrm{s},p,k}$, and $\tilde{c}_{\mathrm{ss},p,k}$ denote the normalized average and surface concentrations of the negative and positive electrodes estimated by the EKF, respectively. $\mathrm{SOC}_{\mathrm{EKF},k}$ is the EKF-updated SOC estimate, $V_{\mathrm{model,pre},k}$ is the model-predicted terminal voltage before the EKF update, and $\nu_k$ is the voltage innovation, defined as

$$\nu_k = V_{\text{meas},k} - V_{\text{model,pre},k} \tag{14}$$

Here, $I_k$ and $T_k$ denote the measured current and temperature, respectively. All input features are standardized using $z$-score normalization based on the training set, and the voltage innovation is clipped to $[-0.5, 0.5]$ V to mitigate transient spikes.

**Table 4.** Input features of the physics-guided GRU residual model.

| Index | Feature | Description |
|---|---|---|
| 1 | $\tilde{\bar{c}}_{s,n}$ | Negative electrode average SOC state (EKF output) |
| 2 | $\tilde{c}_{ss,n}$ | Negative electrode surface SOC state (EKF output) |
| 3 | $\tilde{\bar{c}}_{s,p}$ | Positive electrode average SOC state (EKF output) |
| 4 | $\tilde{c}_{ss,p}$ | Positive electrode surface SOC state (EKF output) |
| 5 | $\text{SOC}_{\text{EKF}}$ | EKF-updated SOC estimate (0%–100%) |
| 6 | $V_{\text{model,pre}}$ | Model-predicted voltage before EKF update (V) |
| 7 | $\nu_k$ | Voltage innovation = $V_{\text{meas},k} - V_{\text{model,pre},k}$ (V) |
| 8 | $I$ | Measured current (A, sign preserved) |
| 9 | $T$ | Measured temperature (°C) |

The corrected SOC estimate is obtained by adding the GRU-predicted residual to the EKF estimate

$$\text{SOC}_{\text{final}}(t) = \text{SOC}_{\text{EKF}}(t) + \alpha \hat{r}(t) \tag{15}$$

where $\hat{r}(t)$ is the GRU-predicted residual, and $\alpha$ is the fusion gain, which is set to 1.0 in this work. The corrected SOC is constrained to the physically meaningful range of 0%–100%. For practical BMS implementation, the residual correction may be further bounded according to admissible SOC limits and safety constraints, while the EKF remains the primary recursive estimation channel.

The GRU is trained to learn the EKF residual, defined as

$$r(t) = \text{SOC}_{\text{ref}}(t) - \text{SOC}_{\text{EKF}}(t) \tag{16}$$

where $SOC_{ref}$ is obtained from Coulomb counting with known initial conditions. Key hyperparameters and configuration of the physics-guided GRU residual learner are shown in Table 5.

**Table 5.** Key hyperparameters and configuration of the physics-guided GRU residual learner.

| **Component** | **Configuration/value** |
|---|---|
| Input features | 9 (electrochemical, EKF, and measurement-based features) |
| Window length/stride | 60 steps/30 steps (50% overlap), yielding a total of 11,277 training windows |
| Hidden layer type | Single-layer GRU |
| Hidden units | 32 |
| Output dimension | 1 (residual correction) |
| Activation | tanh |
| Optimizer | Adam |
| Learning rate/weight decay | $1\times10^{-3}/1\times10^{-5}$ |
| Batch size | 64 |
| Loss function | Mean squared error (MSE) |
| Gradient clipping | max_norm = 1.0 |
| Early stopping / max epochs | 8 epochs patience/50 epochs maximum |
| Training data | DST and FUDS cycles; −10–50 °C; initial SOC errors 0%, 10%, and 20% |
| Test data | US06 cycle (held-out); same temperature and SOC conditions |

The GRU network is implemented as a single recurrent layer with 32 hidden units followed by a linear projection layer that outputs one residual value per time step. A temporal context window of 60 s (stride = 30 s, 50% overlap) is used to provide sufficient temporal correlation while maintaining efficient training. This windowing strategy generates a total of 11,277 training windows. The model is trained with the Adam optimizer (learning rate $1\times10^{-3}$, weight decay $1\times10^{-5}$) using a batch size of 64 and gradient clipping (max_norm = 1.0) to ensure stable convergence. The loss function is the mean-squared error between the predicted and true residuals

$$\mathcal{L} = \frac{1}{N}\sum_{i=1}^{N}[\hat{r}(t_i) - r(t_i)]^2 \tag{17}$$

where $N$ is the number of samples in a batch. Training is terminated early if validation performance does not improve for eight consecutive epochs (maximum 50 epochs).

The training data comprise the DST and FUDS driving cycles at eight temperatures (from −10 to 50 °C) and three initial SOC errors (0%, 10%, and 20%) to improve robustness to initialization uncertainty. The temporal windows overlap only within the training-cycle data used for model fitting. For validation during training, an 80/20 block-level split is applied within the training cycles, and a temporal gap equal to one window length is inserted between training and validation blocks to reduce direct temporal leakage. The US06 cycle is fully held out from training and used as an unseen cross-profile test set under the same temperature and initialization configurations. The reference SOC is obtained by Coulomb counting with known initial conditions, which is standard for this dataset but should be interpreted as a controlled reference trajectory rather than an independently measured internal electrochemical state.

This physics-guided residual learning approach effectively combines the interpretability of electrochemical models with the adaptability of recurrent neural networks. By providing physically meaningful inputs from the EKF, the GRU learns systematic residual patterns that conventional model-based filters cannot capture, leading to more accurate and stable SOC estimation across dynamic operating conditions and wide temperature ranges.

### *2.4 Effect of physics information on PRKL*

This supplementary analysis quantifies the collective contribution of electrochemical state information to the proposed PRKL framework. We compared PRKL with an ablated variant that excludes all electrode-level electrochemical state features, referred to as PRKL-NP (no physics). The ablation is designed as a group-level test of the physical-state channel, rather than a feature-ranking analysis of individual state variables. This choice is appropriate because the average and surface lithiation states of the two electrodes are dynamically coupled through the SPM diffusion equations and should not be interpreted as independent explanatory variables. The comparison was conducted under identical training and evaluation settings across eight temperatures (from −10 to 50 °C), three drive-cycle profiles (DST, US06, and FUDS), and three initialization offsets (0%, 10%, and 20%).

### *2.5 Implementation of cached acceleration technique*

To improve computational efficiency during real-time estimation, a cached acceleration mechanism was introduced to avoid redundant re-discretization of the temperature-dependent state-space model. Discretization of the CPG-SPMT equations (Eqs. (3) and (4)) is computationally expensive, as it must otherwise be repeated whenever temperature-dependent parameters change. The proposed caching strategy stores and reuses previously computed discrete matrices whenever the thermal condition remains within a small tolerance.

Each cache entry is indexed by the fixed sampling interval $\Delta t$ =1 s and contains a tuple {temp, temp_params, $A_d$, $B_d$, $C_d$, $D_d$}, where temp_params represents all temperature-dependent quantities. A cache hit occurs when the current temperature $T$ satisfies

$$|T - T_{\text{entry}}| < 0.1\ ^{\circ}\text{C} \tag{18}$$

and the same $\Delta t$ is used, allowing direct retrieval of the corresponding discrete matrices $(A_{\text{d}}, B_{\text{d}}, C_{\text{d}}, D_{\text{d}})$ and temperature parameters. Otherwise (cache miss), the temperature-dependent parameters are recomputed, and the continuous model is re-discretized using a zero-order hold (ZOH) method to obtain updated matrices $(A_{\text{d}}, B_{\text{d}}, C_{\text{d}}, D_{\text{d}})$, which are then stored for future reuse.

During each time step, the cached matrices are consistently applied in the EKF prediction, measurement, and GRU feature extraction processes to ensure numerical coherence across the hybrid PRKL framework. By reusing pre-discretized matrices under quasi-steady thermal conditions, the caching mechanism substantially reduces redundant computations and matrix operations.

# 3 Results

## *3.1 LFP battery experimental dataset and computational environment*

The experimental dataset used for model development and validation is a public 1.1 Ah 18650 cylindrical graphite/LFP cell dataset reported by the University of Maryland [45]. The dataset comprises 24 experiments covering three dynamic drive cycles, namely dynamic stress test (DST), federal urban driving schedule (FUDS), and supplemental federal test procedure (US06) at eight ambient temperatures: −10, 0, 10, 20, 25, 30, 40, and 50 °C (see Fig. 1). A key advantage of this dataset is that all tests are performed under dynamic load profiles across a wide temperature range, including −10 °C. In contrast to datasets collected only under constant-current profiles or within narrow temperature ranges, this dataset provides a controlled benchmark for evaluating temperature and drive-cycle robustness of SOC estimation methods for the studied LFP cell. For the GRU residual learner, DST and FUDS cycles were used for training at all eight temperatures and three initial SOC errors (0%, 10%, and 20%), while the US06 cycle under the same temperature and initialization settings was used as a held-out cross-profile test set.

All codes and training experiments in this study were executed on a Linux-based workstation equipped with an Intel Xeon Gold 5120 CPU operating at 2.20 GHz, featuring 8 virtual cores and 16 GB of system memory, and an NVIDIA RTX 2080 Ti GPU with 11 GB of VRAM. The entire PRKL framework, including the electrochemical model, EKF implementation, GRU residual learning module, and caching acceleration mechanism, was developed in Python using standard scientific computing libraries (NumPy, SciPy, PyTorch, and Pandas). All results reported in this paper were obtained under this environment to ensure computational consistency and reproducibility.

### *3.2 SPM and parameter identification*

The SPM adopted in this study is based on CPG-SPMT [18,43], which we introduced in Section 2.2. A key advantage of this model is that it has been reformulated into a state-space representation and its observability has been analytically demonstrated [18], which makes it particularly suitable for the design of EKFs. For parameter identification, particle swarm optimization (PSO) was employed due to its stability and accuracy in handling high-dimensional nonlinear problems. The identification was performed using the comprehensive LFP battery dataset introduced in the previous section. The performance of the identified model is illustrated in Fig. 2(a, b), where the model achieves an average root mean square error (RMSE) of 0.033 V, a mean absolute error (MAE) of 0.022 V, and an $R^2$ value of 0.9683. These results indicate that the reduced electrochemical model provides a sufficiently accurate and computationally tractable backbone for recursive SOC estimation within the studied dataset, while remaining imperfect enough to motivate residual correction.

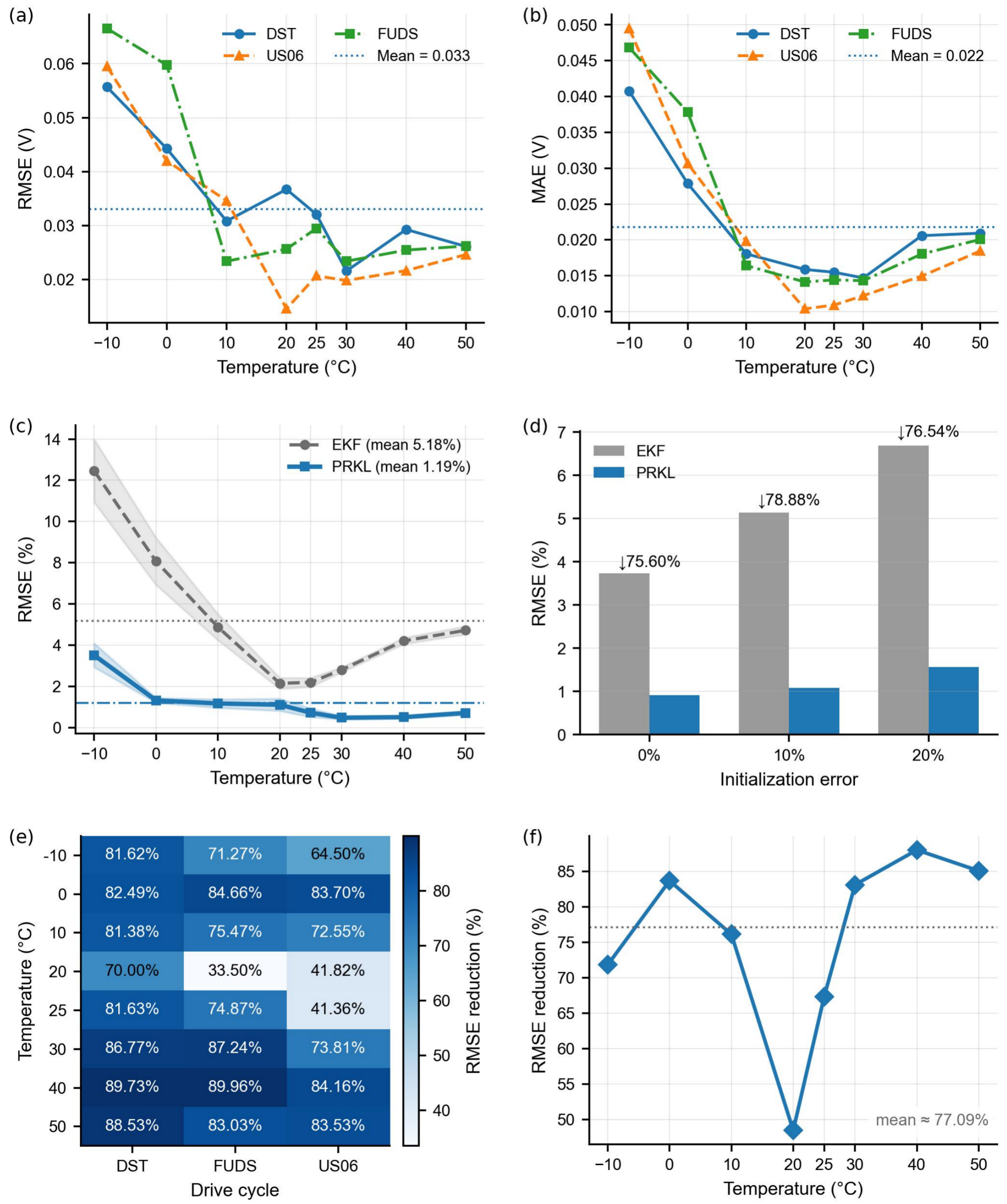

**Fig. 2.** CPG-SPMT model validation and PRKL-enhanced SOC estimation performance. (a, b) Model validation: RMSE (a) and MAE (b) of terminal voltage predictions across temperatures (from −10 to 50 °C) and drive cycles (DST, US06, and FUDS). Dashed lines indicate mean values across all conditions. (c–f) SOC estimation comparison: (c) Temperature-dependent RMSE (%) for EKF (gray) and PRKL (blue); shaded areas represent standard error of the mean (SEM). Horizontal dashed and dash-dotted lines denote global mean RMSE values. (d) Robustness to SOC initialization errors (0%, 10%, and 20%); annotations show relative RMSE reduction by PRKL over EKF. (e) Heatmap of RMSE reduction (%) by PRKL relative to EKF across drive cycles and temperatures; darker blue indicates greater improvement. (f) Temperature-dependent RMSE reduction (%) with global mean (dotted line) indicating overall performance enhancement.

### *3.3 Evaluation of SOC estimation accuracy*

To evaluate the accuracy of the proposed PRKL method for SOC estimation, validation was carried out on three dynamic driving cycles across eight temperatures, using the LFP dataset detailed in Section 3.1. In total, 72 operating conditions were evaluated, covering 8 temperatures, 3 dynamic drive cycles, and 3 initial SOC errors. The results were compared against those obtained with an EKF designed on the basis of the electrochemical model. For consistency, the EKF employed in the baseline and in the PRKL framework was tuned with identical parameters. Fig. 2 and Table 6 together provide a multi-level assessment of SOC estimation accuracy. From a methodological perspective, the results highlight that PRKL substantially improves both accuracy and robustness compared to the conventional EKF. Specifically, Fig. 2(c) shows that the global mean RMSE is reduced from 5.18% for EKF to 1.19%, indicating that the proposed approach achieves a fourfold improvement in accuracy. The temperature-dependent EKF SOC error in Fig. 2(c) generally follows the voltage model error trends in Fig. 2(a, b), suggesting that model inaccuracies substantially influence the SOC estimation performance of the model-based EKF. Within the moderate temperature range from 10 to 50 °C, the EKF achieves an average RMSE below 5%. However, its performance deteriorates significantly under extreme conditions at −10 and 0 °C. The proposed PRKL method maintains lower estimation error even under extreme thermal conditions of −10 and 0 °C. Fig. 2(d) demonstrates that this advantage persists under challenging initialization conditions: even with 20% initial SOC mismatch, PRKL maintains an average RMSE of only 1.57%, a reduction of more than 76% relative to EKF. Fig. 2(e) further shows that the gains are consistent across dynamic drive cycles, with the largest benefit observed for DST (83.26% reduction). These results indicate that PRKL improves SOC estimation accuracy across different operating conditions. Fig. 2(f) confirms that these improvements are robust across the full temperature range, yielding a global mean reduction of 77.1% RMSE.

Table 6 complements these temperature-resolved results by summarizing the average RMSE values across initialization errors and drive cycles, thus providing an aggregated view of robustness of PRKL. On the held-out US06 test cycle, the proposed algorithm attained an average RMSE of 1.61%, indicating cross-profile robustness within the same public LFP dataset. Taken together, Fig. 2 and Table 6 show that PRKL reduces absolute estimation errors and improves resilience to initialization uncertainty, temperature variation, and load dynamics under the present controlled evaluation protocol. These results support the proposed residual-learning mechanism, while broader claims about real-world generalization require validation on additional cells, ageing states, and pack-level data.

**Table 6.** Comprehensive comparison between the EKF and the PRKL method. RMSE values are in %. Decrease (p.p.) = EKF−PRKL; Reduction (%) = (EKF−PRKL)/EKF×100%. Convergence time denotes the time required for SOC estimation to fall within a 5% error range. "N/A" indicates not applicable, and "–" indicates non-convergent cases.

| Group/condition | EKF RMSE (%) | PRKL RMSE (%) | Decrease (p.p.) | Reduction (%) | EKF Conv. (s) | PRKL Conv. (s) | Conditions |
|---|---|---|---|---|---|---|---|
| **Overall performance** | | | | | | | |
| Overall average | 5.18 | 1.19 | 3.99 | 77.03 | N/A | N/A | 72 |
| **Temperature (°C)** | | | | | | | |
| −10 | 12.47 | 3.51 | 8.96 | 71.85 | N/A | N/A | 9 |
| 0 | 8.07 | 1.32 | 6.75 | 83.64 | N/A | N/A | 9 |
| 10 | 4.88 | 1.17 | 3.71 | 76.02 | N/A | N/A | 9 |
| 20 | 2.15 | 1.11 | 1.04 | 48.37 | N/A | N/A | 9 |
| 25 | 2.19 | 0.72 | 1.47 | 67.12 | N/A | N/A | 9 |
| 30 | 2.80 | 0.47 | 2.33 | 83.21 | N/A | N/A | 9 |
| 40 | 4.20 | 0.50 | 3.70 | 88.10 | N/A | N/A | 9 |
| 50 | 4.72 | 0.70 | 4.02 | 85.17 | N/A | N/A | 9 |
| **Drive cycle** | | | | | | | |
| DST | 5.03 | 0.84 | 4.19 | 83.30 | N/A | N/A | 24 |
| FUDS | 4.89 | 1.11 | 3.78 | 77.30 | N/A | N/A | 24 |
| US06 | 5.64 | 1.61 | 4.03 | 71.45 | N/A | N/A | 24 |
| **Initial error** | | | | | | | |
| 0% | 3.74 | 0.91 | 2.83 | 75.67 | N/A | N/A | 24 |
| 10% | 5.14 | 1.08 | 4.06 | 78.99 | N/A | N/A | 24 |
| 20% | 6.68 | 1.57 | 5.11 | 76.50 | N/A | N/A | 24 |
| **Examples** | | | | | | | |
| DST, −10 °C, 0% | 7.86 | 1.52 | 6.34 | 80.66 | – | – | 1 |
| DST, −10 °C, 10% | 10.94 | 1.96 | 8.98 | 82.08 | – | 1 | 1 |
| DST, −10 °C, 20% | 16.96 | 3.09 | 13.87 | 81.78 | – | 5 | 1 |
| US06, 30 °C, 0% | 2.23 | 0.28 | 1.95 | 87.44 | – | – | 1 |
| US06, 30 °C, 10% | 2.52 | 0.52 | 2.00 | 79.37 | 29 | 1 | 1 |
| US06, 30 °C, 20% | 2.72 | 1.16 | 1.56 | 57.35 | 132 | 7 | 1 |
| FUDS, 50 °C, 0% | 4.60 | 0.50 | 4.10 | 89.13 | – | – | 1 |
| FUDS, 50 °C, 10% | 6.04 | 1.38 | 4.66 | 77.15 | 15 | 1 | 1 |
| FUDS, 50 °C, 20% | 4.60 | 0.71 | 3.89 | 84.57 | 59 | 6 | 1 |

Several representative cases are discussed here to illustrate the performance of the proposed method in more detail. Fig. 3 shows the SOC estimation and error curves under three representative conditions: DST at −10 °C, US06 at 30 °C, and FUDS at 50 °C. Table 6 summarizes the comparison of SOC estimation accuracy and convergence time (within a 5% error bound) between EKF and PRKL. As shown in both Fig. 3 and Table 6, PRKL consistently outperforms the baseline EKF across all three conditions, demonstrating the superiority of the proposed approach.

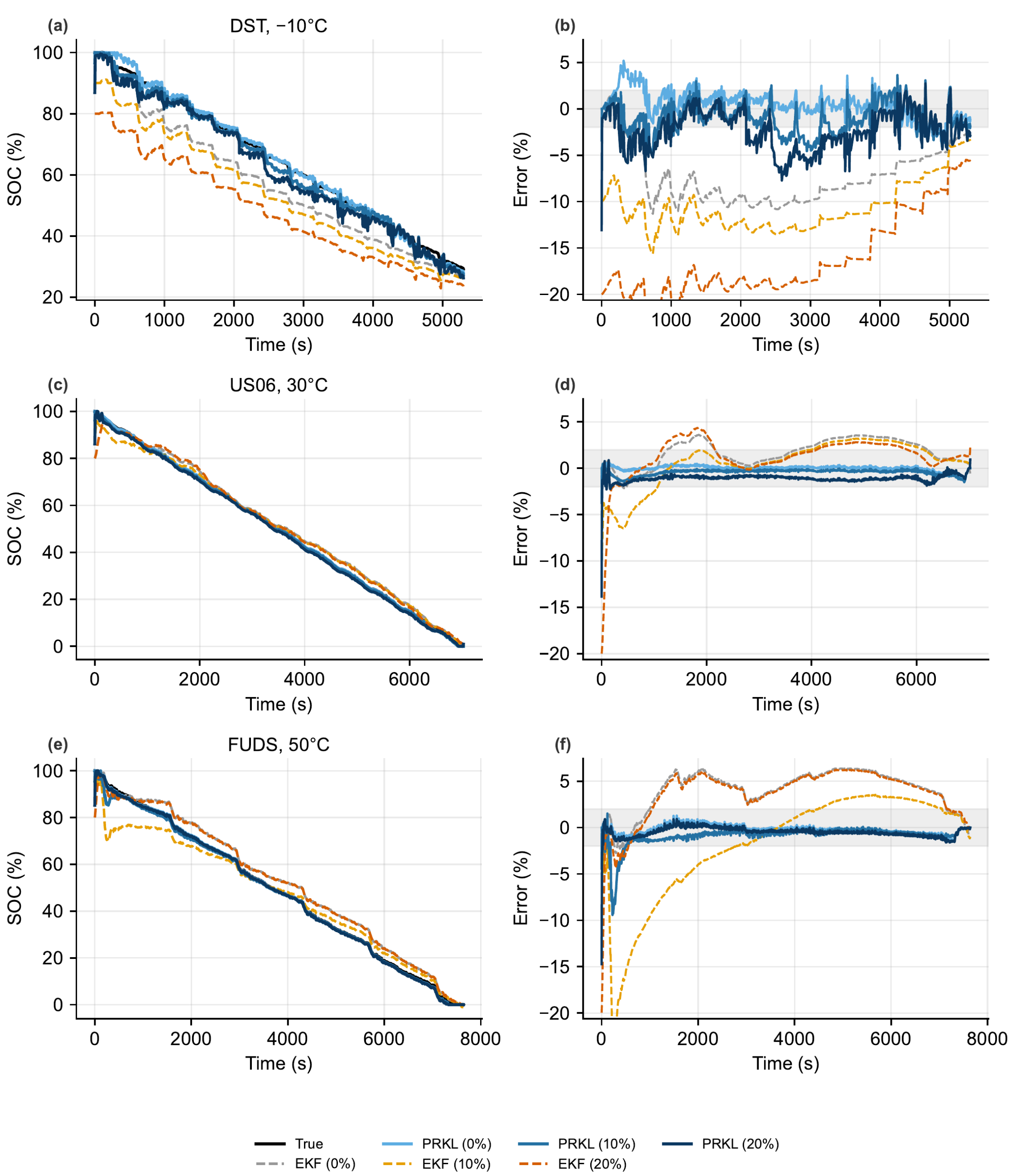

**Fig. 3.** Temporal comparison of SOC estimation across representative operating conditions. Each row corresponds to a distinct drive cycle and temperature: (a, b) DST at −10 °C, (c, d) US06 at 30 °C, and (e, f) FUDS at 50 °C. Left panels show the estimated SOC trajectories by EKF (gray-orange-red dashed lines) and PRKL (light-medium-dark blue solid lines) under initial SOC errors of 0%, 10%, and 20%, compared to the true SOC (black). Right panels depict the corresponding estimation errors, with the shaded regions indicating a ±2% tolerance band.

The EKF exhibits large estimation errors under the −10 °C DST condition and fails to converge under 10% and 20% initialization errors (Fig. 3a). This low-temperature degradation is expected for several coupled reasons. Solid-phase diffusion and charge-transfer kinetics slow down at lower temperatures, ohmic and polarization losses increase, and the sensitivity of the voltage response to uncertain parameters becomes stronger. For LFP cells, the flat OCV-SOC plateau further weakens the voltage information available for SOC correction. These effects amplify model mismatch and reduce the corrective capability of the EKF. Although PRKL converges toward the correct SOC trajectory, its error remains higher at −10 °C compared to moderate temperatures because the residual learner receives electrochemical states generated by a less accurate low-temperature model. The present framework therefore reduces but does not eliminate low-temperature error. A dedicated low-temperature compensation strategy, such as temperature-specific parameter adaptation or uncertainty-weighted residual fusion, is left for future work.

Convergence speed was quantified as the time required for the SOC estimation error to first fall within 5%. Given the sampling interval of 1 s, as summarized in Table 6, under a 10% initialization error, PRKL converges within 1 s, whereas the EKF requires 29 s under the US06 (30 °C) condition and 15 s under the FUDS (50 °C) condition, and fails to converge under the DST (−10 °C) condition. Even with a 20% initialization error, PRKL requires only 5–7 s to reach the 5% accuracy threshold, while the EKF needs 132 s (US06, 30 °C) and 59 s (FUDS, 50 °C), with no convergence again at DST (−10 °C).

### *3.4 Effect of physics information on PRKL*

To highlight the benefits of the electrochemical information input to the GRU residual learning network, a control variant, denoted as PRKL-NP (NP = No Physics), was constructed without physics-based input features. Unlike the original PRKL, PRKL-NP excludes the four electrochemical states ($\tilde{\bar{c}}_{\mathrm{s},n}$, $\tilde{c}_{\mathrm{ss},n}$, $\tilde{\bar{c}}_{\mathrm{s},p}$, and $\tilde{c}_{\mathrm{ss},p}$) from the input feature set, while retaining the remaining five features: $\mathrm{SOC}_{\mathrm{EKF}}$, $V_{\mathrm{model,pre}}$, Innovation, Current, and Temperature. The training procedure and hyperparameter settings of PRKL-NP were kept identical to those of PRKL to ensure a fair comparison. The results are shown in Fig. 4 and Table 7. Both PRKL and PRKL-NP exhibit lower error distributions and medians compared to the EKF baseline, and PRKL achieves the smallest overall error spread (Fig. 4a). Quantitatively, PRKL reduces the average RMSE by 47.81% relative to PRKL-NP across all test conditions. The empirical cumulative distribution functions (ECDFs) in Fig. 4(b) further illustrate this improvement, where the PRKL curve consistently shifts leftward, indicating smaller estimation errors than PRKL-NP. These results demonstrate that the prediction power of the model declines when the electrochemical information is excluded from the input features to the residual learning network.

Fig. 4(c and d) compare PRKL and PRKL-NP under different temperatures and drive cycles, respectively. In all cases, PRKL outperforms PRKL-NP, confirming the added value of the physics-based features. The smallest improvement is observed at −10 °C, likely because the electrochemical model incurs higher errors under extremely low temperatures, which weakens the added value of the physical information. Fig. 4(e) further shows that PRKL consistently outperforms PRKL-NP across all drive cycles, while Fig. 4(f) reveals that incorporating physical information also enhances robustness against initialization errors.

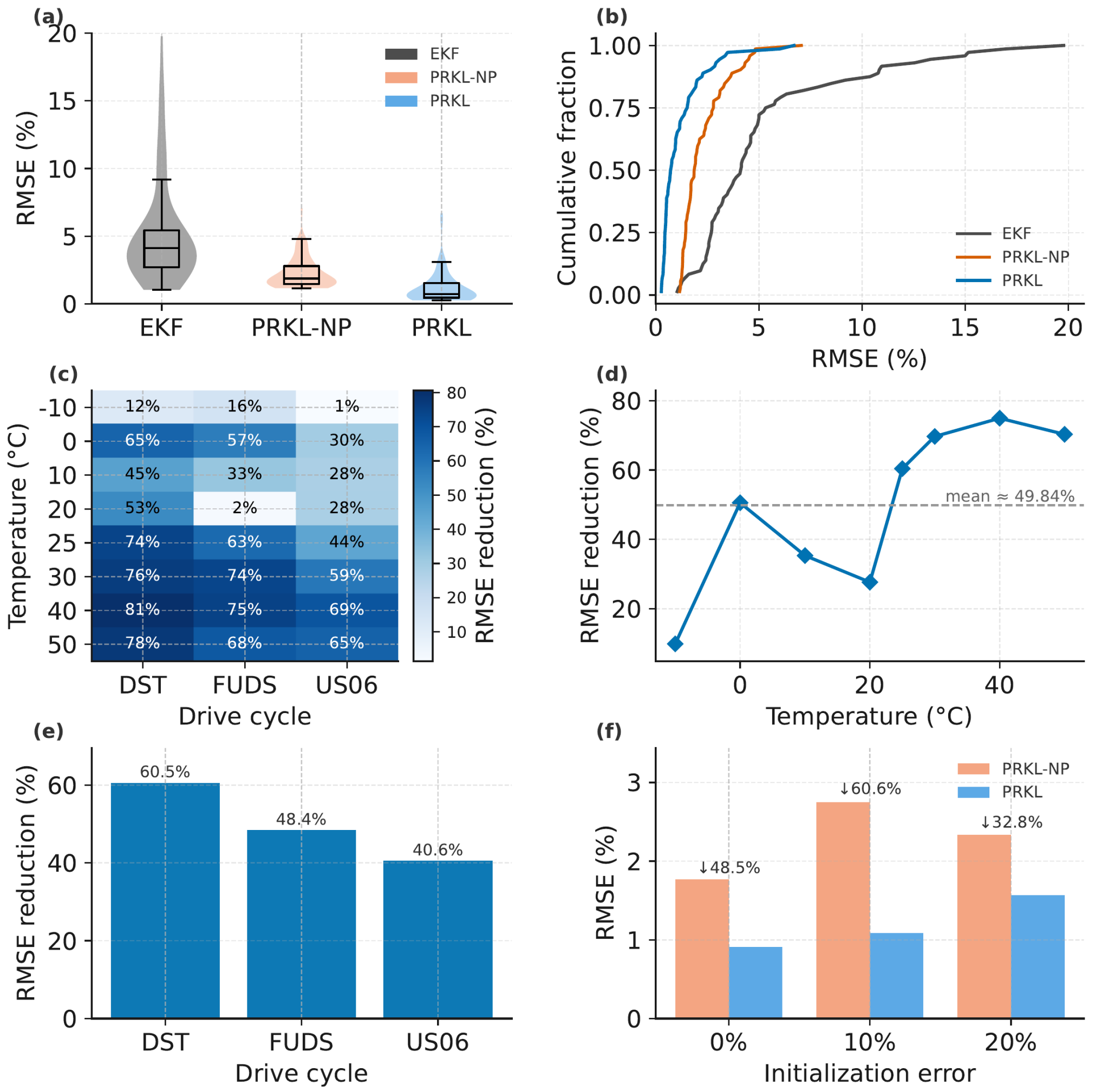

**Fig. 4.** Effect of physics information on PRKL for battery SOC estimation. (a) Violin/box plots of absolute SOC RMSE (%) for a classical EKF (grey), an ablated model without physics PRKL-NP (orange), and the physics-guided PRKL (blue). (b) ECDFs of absolute RMSE (%) for the same three methods. (c) Heat map of RMSE reduction (%) of PRKL relative to PRKL-NP across temperature (*y*-axis) and drive cycle (*x*-axis; DST, FUDS, US06); numbers denote mean reductions and darker cells indicate larger gains. (d) Temperature dependence of the RMSE reduction (%) (mean within each temperature; dashed line marks the across-temperature mean). (e) Drive-cycle dependence of the RMSE reduction (%) with mean values above the bars. (f) Absolute RMSE (%) for PRKL-NP and PRKL at 0%, 10%, and 20% initialization error; arrows annotate the corresponding RMSE reduction (%) of PRKL relative to PRKL-NP, computed as (RMSE_PRKL-NP − RMSE_PRKL)/RMSE_PRKL-NP × 100%.

As summarized in Table 7, incorporating physics-based features, including the four electrode-level SOC states, improves estimation accuracy across all conditions. On average, PRKL reduced the SOC RMSE from 2.28% (PRKL-NP) to 1.19%, corresponding to a mean absolute decrease of 1.10% points and a relative reduction of 47.81%. The improvement is the most pronounced at mid-to-high temperatures (25–50 °C), where the reduction exceeds 60%–70%, suggesting that the model effectively leverages temperature-dependent electrochemical dynamics.

Performance gains are also consistent across the three driving profiles in the studied dataset, with the largest improvement observed for the DST cycle (60.48% reduction) and a notable 48.43% reduction for FUDS. Furthermore, PRKL demonstrates stronger robustness to initialization uncertainty, achieving 63.6% lower RMSE when the EKF is initialized with a 10% SOC offset. These results indicate that embedding electrochemical state information improves residual correction at the group-feature level. They do not imply that each electrochemical input feature has an independent contribution weight, because the electrode average and surface states are physically coupled through the diffusion model.

**Table 7.** Comparison between PRKL and its ablated counterpart without physics information (PRKL-NP). Values are means over matched conditions. RMSE values are in %. Decrease (p.p.) is PRKL-NP−PRKL in percentage points; Reduction (%) is (RMSEPRKL-NP−RMSEPRKL)/RMSEPRKL-NP×100%.

| Group | PRKL-NP RMSE (%) | PRKL RMSE (%) | Decrease (p.p.) | Reduction (%) | Conditions |
|---|---|---|---|---|---|
| **Overall** | | | | | |
| Overall | 2.28 | 1.19 | 1.10 | 47.81 | 72 |
| **Temperature (°C)** | | | | | |
| −10 | 3.98 | 3.51 | 0.47 | 11.81 | 9 |
| 0 | 2.90 | 1.32 | 1.59 | 54.48 | 9 |
| 10 | 1.78 | 1.17 | 0.61 | 34.27 | 9 |
| 20 | 1.56 | 1.11 | 0.46 | 28.85 | 9 |
| 25 | 1.81 | 0.72 | 1.09 | 60.22 | 9 |
| 30 | 1.60 | 0.47 | 1.13 | 70.62 | 9 |
| 40 | 2.20 | 0.50 | 1.69 | 77.27 | 9 |
| 50 | 2.44 | 0.70 | 1.74 | 71.31 | 9 |
| **Cycle** | | | | | |
| DST | 2.26 | 0.84 | 1.42 | 62.83 | 24 |
| FUDS | 2.11 | 1.11 | 1.00 | 47.39 | 24 |
| US06 | 2.49 | 1.61 | 0.88 | 35.34 | 24 |
| **Initial error** | | | | | |
| 0% | 1.77 | 0.91 | 0.86 | 48.59 | 24 |
| 10% | 2.75 | 1.08 | 1.67 | 60.73 | 24 |
| 20% | 2.33 | 1.57 | 0.77 | 32.62 | 24 |

*3.5 Runtime acceleration through cached optimization*

To further enhance computational efficiency, a caching-based acceleration strategy was introduced into the PRKL framework. When the temperature difference exceeds this threshold, new temperature-dependent parameters are computed, and the discrete matrices are updated accordingly. This approach eliminates redundant discretization operations while maintaining numerical consistency and estimation accuracy. Table 8 summarizes the performance comparison between the baseline PRKL and its cached variant. The results show that the cached PRKL preserves virtually identical estimation accuracy (overall RMSE ~1.19%), while substantially improving computational efficiency. The average per-step runtime decreases from 1.29 to 0.63 ms, corresponding to a 51% reduction in calculation time relative to PRKL and approximately 20% faster than the standard EKF under the workstation environment used in this study. These results confirm that the caching mechanism accelerates computation without compromising accuracy. However, they should be interpreted as evidence of algorithmic computational feasibility rather than proof of completed embedded BMS deployment, because MCU/DSP runtime, RAM usage, flash occupancy, fixed-point effects, and real-time scheduling were not evaluated in this work.

**Table 8.** Accuracy and runtime comparison of PRKL and PRKL_cached, with EKF runtime for reference. RMSE values are in %. Time is the per-step average in ms. Time reductions are computed relative to EKF and non-cached PRKL.

| Group | PRKL RMSE (%) | PRKL_cached RMSE (%) | EKF time (ms) | PRKL time (ms) | PRKL_cached time (ms) | Time reduction vs. EKF (%) | Time reduction vs. PRKL (%) | Conditions |
|---|---|---|---|---|---|---|---|---|
| **Overall** | | | | | | | | |
| Overall | 1.1884 | 1.1887 | 0.78 | 1.29 | 0.63 | 19.23 | 51.16 | 72 |
| **Temperature (°C)** | | | | | | | | |
| −10 | 3.5170 | 3.5261 | 0.78 | 1.29 | 0.63 | 19.23 | 51.16 | 9 |
| 0 | 1.3190 | 1.3180 | 0.78 | 1.29 | 0.63 | 19.23 | 51.16 | 9 |
| 10 | 1.1656 | 1.1641 | 0.78 | 1.27 | 0.62 | 20.51 | 51.18 | 9 |
| 20 | 1.1060 | 1.1035 | 0.78 | 1.29 | 0.63 | 19.23 | 51.16 | 9 |
| 25 | 0.7170 | 0.7162 | 0.79 | 1.28 | 0.63 | 20.25 | 50.78 | 9 |
| 30 | 0.4740 | 0.4747 | 0.78 | 1.29 | 0.62 | 20.51 | 51.94 | 9 |
| 40 | 0.5037 | 0.5015 | 0.78 | 1.29 | 0.63 | 19.23 | 51.16 | 9 |
| 50 | 0.7045 | 0.7054 | 0.78 | 1.29 | 0.63 | 19.23 | 51.16 | 9 |
| **Cycle** | | | | | | | | |
| DST | 0.8425 | 0.8555 | 0.78 | 1.28 | 0.63 | 19.23 | 50.78 | 24 |
| FUDS | 1.1080 | 1.1035 | 0.78 | 1.29 | 0.63 | 19.23 | 51.16 | 24 |
| US06 | 1.6145 | 1.6070 | 0.79 | 1.29 | 0.62 | 21.52 | 51.94 | 24 |
| **Initial SOC** | | | | | | | | |
| 0% | 0.9131 | 0.9136 | 0.78 | 1.29 | 0.62 | 20.51 | 51.94 | 24 |
| 10% | 1.0851 | 1.0931 | 0.79 | 1.28 | 0.63 | 20.25 | 50.78 | 24 |
| 20% | 1.5668 | 1.5594 | 0.78 | 1.29 | 0.63 | 19.23 | 51.16 | 24 |

# 4 Discussion

This study introduced PRKL as a mechanism-oriented framework for improving electrochemical-model-based recursive SOC estimation of LFP batteries. The main result is that the residual errors of an SPM-based EKF can be substantially reduced when the residual learner receives physically meaningful electrochemical states in addition to measurement-level variables. Across 72 operating conditions in the studied public LFP dataset, PRKL reduced the average SOC RMSE from 5.18% for the EKF to 1.19%, while the no-physics ablation (PRKL-NP) confirmed the added value of the electrochemical state channel. The work should therefore be interpreted as evidence for physics-guided residual correction of an electrochemical estimator, rather than as an exhaustive benchmark of all possible SOC estimation algorithms.

The roles of the three main components are complementary. The CPG-SPMT model provides a low-order electrochemical backbone with interpretable electrode-level states and acceptable computational cost. A higher-fidelity P2D model could describe more internal phenomena, but it would introduce a high-dimensional PDE-constrained state space, substantially higher parameter-identification burden, and much greater computational cost for recursive filtering. Replacing EKF with UKF may reduce local linearization error, but it would not directly remove systematic residuals caused by model simplification, parameter mismatch, temperature-dependent polarization, and the weak voltage-SOC sensitivity of LFP cells. The GRU is used as a lightweight temporal residual learner; the novelty is not the GRU architecture itself, but the decomposition into physics-based state propagation and learned residual correction. Other sequence learners, such as LSTM, TCN, or Transformer modules, could in principle be placed in the residual channel, but architecture benchmarking is outside the scope of this work.

The validation protocol also defines the limits of the generalization claim. The US06 cycle is fully held out from training and therefore supports cross-profile testing within the same cell, chemistry, and test campaign. It does not demonstrate broad generalization across manufacturers, capacities, ageing states, cell formats, or pack-level systems. The use of Coulomb-counting-derived SOC with known initial conditions is standard for this public dataset, but it is still a controlled reference trajectory rather than an independent measurement of internal lithium inventory. The overlapping windows used for GRU training increase temporal samples, while the block-level validation split and held-out drive cycle reduce leakage risk. Nevertheless, deployment-level generalization requires external datasets and cell-to-cell validation.

The low-temperature results clarify an important boundary of the current framework. At −10 °C, PRKL still significantly outperforms the EKF, but the RMSE increases to 3.51%. This behavior is consistent with slower diffusion, larger polarization, stronger parameter sensitivity, and weaker voltage observability in the flat LFP OCV region. Because the residual learner is guided by EKF-derived electrochemical states, low-temperature model mismatch can reduce the reliability of the physical features supplied to the GRU. This suggests that at lower temperatures, additional mechanisms are involved, which may require the inclusion of temperature-specific parameter adaptation, uncertainty-weighted residual fusion, or low-temperature-focused training data.

Several practical aspects remain open. The cached implementation reduced the workstation per-step runtime from 1.29 to 0.63 ms without changing accuracy, indicating that redundant model discretization can be avoided efficiently. However, this result is not a substitute for embedded validation on a representative BMS microcontroller or DSP, where memory footprint, flash occupancy, scheduling, and numerical precision must be assessed. In addition, the present PRKL implementation is deterministic and does not quantify the uncertainty of the neural residual. In safety-critical BMS applications, future implementations should combine bounded residual correction, confidence-aware fusion, and formal robustness analysis. The EKF channel provides a physically interpretable baseline estimate, and the residual correction can be constrained in practice, but a formal stability proof for the complete nonlinear EKF-plus-neural-residual system remains for future work.

Overall, PRKL demonstrates that electrochemical model states can provide useful structure for residual learning in LFP SOC estimation. The framework improves accuracy and convergence under the tested drive cycles, temperatures, and initialization offsets, while preserving a clear separation between physical state propagation and data-driven correction. Future work should extend the validation to aged cells, different LFP manufacturers and capacities, other chemistries, and pack-level data, and should integrate uncertainty-aware residual fusion and embedded hardware testing.

## 5 Conclusions

This work proposed a physics-guided residual Kalman learning framework for SOC estimation of graphite/LFP cells. PRKL preserves an electrochemical SPM-based EKF as the recursive estimation backbone and adds a GRU residual correction channel that uses electrochemical states, voltage innovation, current, temperature, and EKF outputs to compensate structured estimation errors.

Within the studied public LFP dataset, covering three dynamic drive cycles, eight temperatures (from −10 to 50 °C), and three initialization offsets (0%, 10%, and 20%), PRKL achieved a global mean SOC RMSE of 1.19%, compared with 5.18% for the physics-only EKF. On the held-out US06 cycle, the mean RMSE was 1.61%, supporting cross-profile robustness within the same cell dataset. The ablation against PRKL-NP further showed that electrode-level electrochemical state features improve residual learning, reducing RMSE by 47.81% relative to the no-physics residual-learning variant.

A cached discretization strategy reduced workstation per-step runtime from 1.29 to 0.63 ms, while preserving the same estimation accuracy, indicating that the proposed algorithm can be optimized for real-time-oriented computation. This result should be viewed as a step toward embedded implementation, rather than a completed MCU/DSP deployment. Further work is needed to validate PRKL across different cells, manufacturers, ageing states, and pack-level systems, and to incorporate uncertainty-aware fusion, bounded residual correction, and formal robustness analysis for safety-critical BMS applications.

**CRediT authorship contribution statement**

Feng Guo: Conceptualization, Methodology, Software, Formal analysis, Investigation, Data curation,

Visualization, Funding acquisition, Writing –original draft. Luis D. Couto: Methodology, Writing – review & editing. Khiem Trad: Investigation, Data curation, Writing – review & editing. Ru Hong: Validation, Resources, Writing – review & editing. Guangdi Hu: Supervision, Writing – review & editing. Mohammadhosein Safari: Conceptualization, Supervision, Writing – review & editing.

**Acknowledgments**

This work was supported by the Research Foundation - Flanders (FWO) (grant numbers 1252326N).

**Data availability**

All data and model used in this study are publicly available on the GitHub repository: https://github.com/FrankSuperG/CPG-SPMT.

**References**

[1] J. Zhao, X. Han, Y. Wu, Z. Wang, A.F. Burke, J. Energy Chem. 102 (2025) 463–496.
[2] J. Schoberl, M. Ank, M. Schreiber, N. Wassiliadis, M. Lienkamp, eTransportation 19 (2024) 100305.
[3] H. Liu, C. Li, X. Hu, J. Li, K. Zhang, Y. Xie, R. Wu, Z. Song, Nat. Commun. 16 (2025) 1137.
[4] K. Luo, X. Chen, H. Zheng, Z. Shi, J. Energy Chem. 74 (2022) 159–173.
[5] Z.P. Cano, D. Banham, S. Ye, A. Hintennach, J. Lu, M. Fowler, Z. Chen, Nat. Energy 3 (2018) 279–289.
[6] G. Harper, R. Sommerville, E. Kendrick, L. Driscoll, P. Slater, R. Stolkin, A. Walton, P. Christensen, O. Heidrich, S. Lambert, A. Abbott, K. Ryder, L. Gaines, P. Anderson, Nature 575 (2019) 75–86.
[7] Z. Zhang, Y. Li, H. Wang, L. Lu, X. Han, D. Li, M. Ouyang, J. Energy Storage 75 (2024) 109696.
[8] F. Guo, L.D. Couto, G. Mulder, K. Trad, G. Hu, O. Capron, K. Haghverdi, J. Energy Storage 101 (2024) 113850.
[9] K. Haghverdi, D.L. Danilov, G. Mulder, L.D. Couto, F. Guo, R.-A. Eichel, J. Power Sources 652 (2025) 237483.
[10] L. Ma, T. Zhang, J. Energy Chem. 80 (2023) 48–57.
[11] Y. Che, L. Xu, R. Teodorescu, X. Hu, S. Onori, ACS Energy Lett. 10 (2025) 741–749.
[12] A. Nekahi, K. Zaghib, Energy Storage Mater. 84 (2026) 104759.

[13] M. Safari, C. Delacourt, J. Electrochem. Soc. 158 (2011) A562–A571.
[14] C. Wang, M. Yang, X. Wang, Z. Xiong, F. Qian, C. Deng, C. Yu, Z. Zhang, X. Guo, J. Energy Storage 110 (2025) 115346.
[15] L. Wu, Z. Lyu, Z. Huang, C. Zhang, C. Wei, J. Energy Chem. 89 (2024) 27–40.
[16] T.F. Fuller, M. Doyle, J. Newman, J. Electrochem. Soc. 141 (1994) 1–10.
[17] F. Guo, L.D. Couto, J. Power Sources 650 (2025) 237365.
[18] F. Guo, L.D. Couto, J. Power Sources 649 (2025) 237309.
[19] G.L. Plett, J. Power Sources 134 (2004) 277–292.
[20] Z. Chen, H. Sun, G. Dong, J. Wei, J. Wu, J. Power Sources 414 (2019) 158–166.
[21] B.G. Choobar, H. Hamed, M. Safari, J. Energy Storage 96 (2024) 112628.
[22] W. Qian, W. Li, X. Guo, H. Wang, Energy 292 (2024) 130585.
[23] Q. Zhu, L. Li, X. Hu, N. Xiong, G.-D. Hu, IEEE Trans. Veh. Technol. 66 (2017) 10853–10865.
[24] Y. Xiong, D. Zhang, X. Ruan, S. Jiang, X. Zou, W. Yuan, X. Liu, Y. Zhang, Z. Nie, D. Wei, Y. Zeng, P. Cao, G. Zhang, Energy Storage Mater. 73 (2024) 103860.
[25] X. Ren, S. Liu, X. Yu, X. Dong, Energy 234 (2021) 121236.
[26] M.A. Hannan, D.N. How, M.H. Lipu, M. Mansor, P.J. Ker, Z. Dong, K. Sahari, S.K. Tiong, K.M. Muttaqi, T.I. Mahlia, F. Blaabjerg, Sci. Rep. 11 (2021) 19541.
[27] W. Qi, W. Qin, Z. Yun, Energy 307 (2024) 132805.
[28] K.-H. Kim, K.-H. Oh, H.-S. Ahn, H.-D. Choi, IEEE Trans. Power Electron. 39 (2023) 125–134.
[29] M. Korkmaz, Energy 294 (2024) 130913.
[30] J. Tian, C. Chen, W. Shen, F. Sun, R. Xiong, Energy Storage Mater. 61 (2023) 102883.
[31] W. Ma, Y. Lei, X. Wang, B. Chen, J. Energy Chem. 80 (2023) 768–784.
[32] H. Sorouri, A. Oshnoei, Y. Che, R. Teodorescu, J. Energy Storage 100 (2024) 113604.
[33] D. Raabe, J.R. Mianroodi, J. Neugebauer, Nat. Comput. Sci. 3 (2023) 198–209.
[34] J. Chen, T. Hannan, Y. Yao, G. Song, Energy Storage Mater. 72 (2024) 103730.
[35] H. Zhao, Q. Li, J. Hu, J. Energy Storage 124 (2025) 116841.
[36] O. Rezaei, A. Rahdan, S. Sardari, M. Dahmardeh, Z. Wang, J. Energy Storage 68 (2023) 107883.
[37] F. Feng, S. Teng, K. Liu, J. Xie, Y. Xie, B. Liu, K. Li, J. Power Sources 455 (2020) 227935.
[38] Q. Zheng, X. Yin, D. Zhang, J. Energy Storage 73 (2023) 109244.
[39] J. Tian, R. Xiong, J. Lu, C. Chen, W. Shen, Energy Storage Mater. 50 (2022) 718–729.
[40] Y. Zeng, Y. Li, T. Yang, Batteries 9 (2023) 358.
[41] J. Tian, R. Xiong, W. Shen, J. Lu, Appl. Energy 291 (2021) 116812.
[42] J. Hou, J. Xu, C. Lin, D. Jiang, X. Mei, Energy 290 (2024) 130056.
[43] S. Zhang, M. Liu, R. Guo, J. Tian, Z. Man, W. Shen, IEEE Trans. Transp. Electrific. 12 (2026) 1223–1234.
[44] F. Guo, L.D. Couto, Comput. Phys. Commun. 322 (2026) 110075.
[45] CALCE Battery Research Group, A123 LFP 18650 battery dataset, Center for Advanced Life Cycle Engineering, University of Maryland, https://calce.umd.edu/battery-data/#A123, accessed 10 February 2026.

## Graphical Abstract

Electrochemical state propagation and gated recurrent unit (GRU) residual learning are integrated to correct state of charge (SOC) errors of the extended Kalman filter (EKF), improving lithium iron phosphate (LFP) battery estimation under dynamic cycles, temperature variation, and initialization uncertainty.

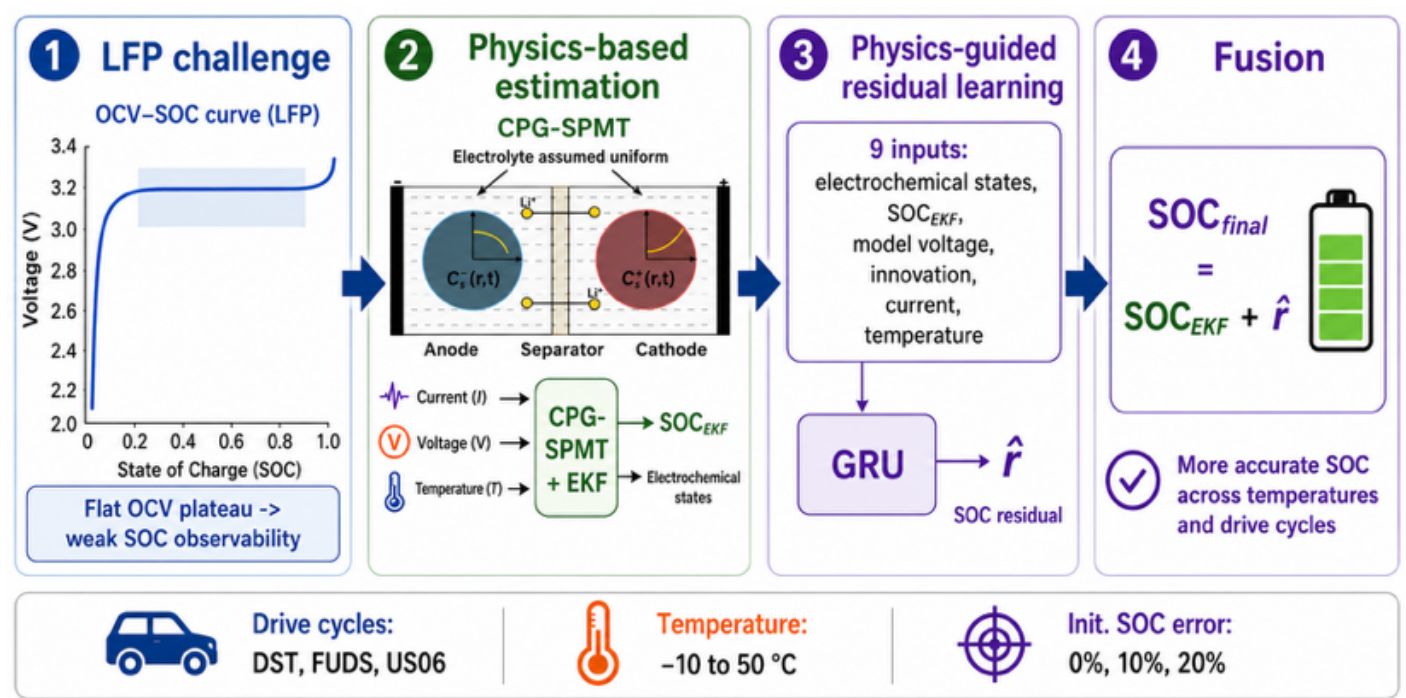

### Biography of the main authors

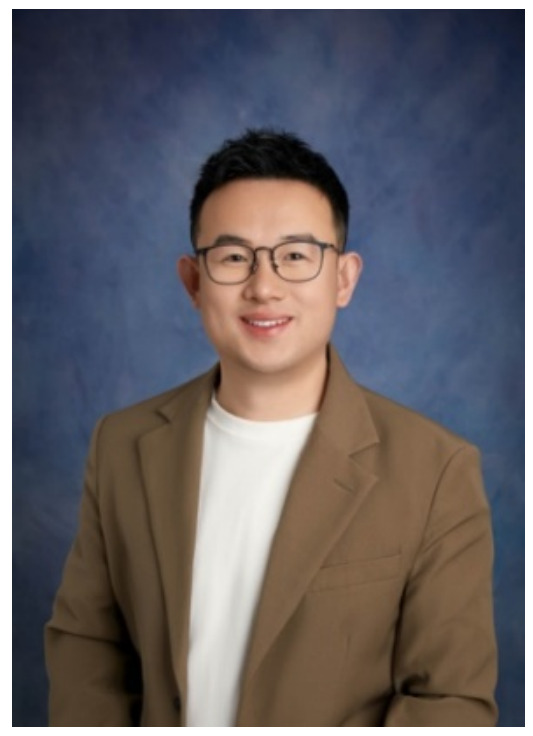

**Feng Guo** is currently an FWO Senior Postdoctoral Fellow at Hasselt University and VITO (Flemish Institute for Technological Research), Belgium. He received his Bachelor's degree in Mechanical Engineering from Southwest Jiaotong University, Chengdu, China, in 2015, and his Ph.D. degree in Automotive Engineering from the same university in 2020, where his doctoral research focused on battery state estimation for electric vehicles. After gaining experience in the industry, he joined VITO in 2022 as a Postdoctoral Researcher. In October 2025, he started his FWO-funded Senior Postdoctoral Fellowship at Hasselt University, where his research focuses on electrochemical modeling, battery state estimation, fault diagnosis, control, and physics-guided learning.

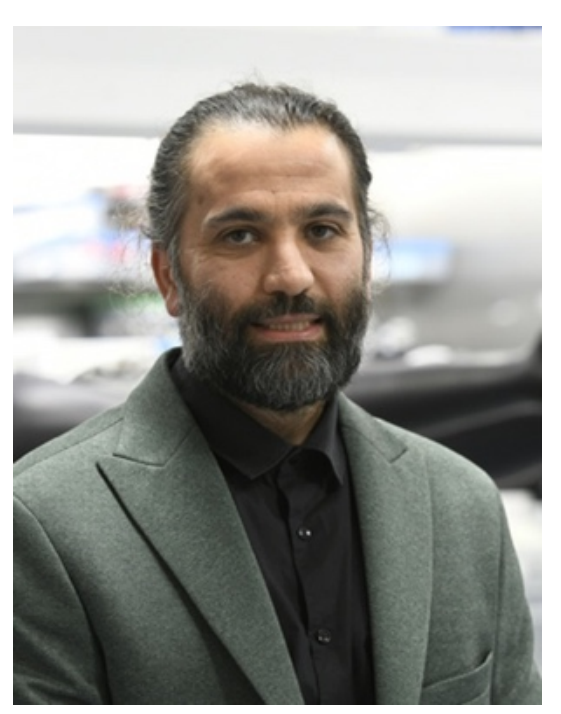

**Mohammadhosein Safari** is a Professor at faculty of Engineering Technology of Hasselt University and leads the group of Electrochemical Engineering (EE) at the Institute for the Materials Research (IUMAT) and EnergyVille. He earned his PhD degree in Electrochemical Engineering from Université de Picardie Jules Verne (France) in 2012, working with Dr. Charles Delacourt and Prof. Jean-Marie Tarascon. Momo was a postdoctoral Fellow in the groups of Prof. Linda Nazar and Prof. Michael Fowler at University of Waterloo (Canada) before joining the Hasselt University in 2015. Momo is interested in the fundamental investigation of electrochemical systems to shed light on and quantify the charge transfer reactions and charge transport phenomena in the electrodes and electrolytes, as well as the aging phenomena.